\documentclass[%
reprint,
superscriptaddress,r
 amsmath,amssymb,
 aps,
pra,
]{revtex4-2}

\usepackage{graphicx}
\usepackage{dcolumn}
\usepackage{bm}
\usepackage{mathtools}
\usepackage{xcolor}
\usepackage{amsmath}
\usepackage{bm}
\usepackage{subcaption}

\newcommand{\ket}[1]{|#1\rangle}

\newcommand{\braket}[2]{\langle#1|#2\rangle}
\newcommand{\braXket}[3]{\langle#1|#2|#3\rangle}

\newcommand {\fabsq}[1] {\left| #1 \right|^{2}}

\newcommand {\fnorm} [1] {\left\| #1 \right\|}
\newcommand {\fnormq} [1] {{\left\| #1 \right\|}^{2}}

\newcommand {\cG} {{\cal G}}

\newcommand {\cR} {{\cal R}}

\newcommand {\bK}{\boldsymbol{K}}
\newcommand {\blambda}{\boldsymbol{\lambda}}
\newcommand {\bepsilon}{\boldsymbol{\epsilon}}
\newcommand {\bnull}{\boldsymbol{0}}

\begin{document}

\title{Enhanced shortcuts to adiabaticity in an interacting giant-atom–atomic-waveguide system}
\author{Ruijin Sun}
\affiliation{%
 Center for Quantum Science and School of Physics, Northeast Normal University, Changchun 130024,China
}%
\affiliation{%
 School of Physics, University College Cork, Cork, Ireland}%

\author{Zhihai Wang}
\email{wangzh761@nenu.edu.cn}
\affiliation{%
 Center for Quantum Science and School of Physics, Northeast Normal University, Changchun 130024,China
}%

\author{Andreas Ruschhaupt}
\email{aruschhaupt@ucc.ie}
\affiliation{%
School of Physics, University College Cork, Cork, Ireland}%

\begin{abstract}

In realistic quantum systems, isolated few-level models usually provide
only an approximate description, since the controlled subsystem is
generally coupled to additional degrees of freedom. As a result, 
a quantum control scheme
designed for an isolated system may lose
its original performance in a more complex coupled setting. Here, we
consider a giant atom coupled to an interacting atomic waveguide and
investigate enhanced shortcuts to adiabaticity (eSTA) in such
a structured system. 
As a first step, a weak-coupling perturbative analysis identifies how changes in the initial and target states, together with dynamical transitions, contribute to the fidelity loss.
Starting from a shortcut to adiabaticity control scheme for the uncoupled system, the enhanced STA correction improves the final fidelity by modifying only the giant-atom control fields.
This improvement is found for both periodic and open boundary conditions of the atomic waveguide.
The results demonstrate a way to design an eSTA protocol for an interacting composite system.
\end{abstract}

\maketitle

\section{\label{sec:level1}Introduction}

Multi--atom quantum systems provide an important platform for studying
complex quantum dynamics, many-body interactions and quantum information
processing. In particular, neutral atom arrays based on optical tweezers enable the trapping and individual control of single atoms, while offering
good scalability, long coherence times~\cite{manetsch2025tweezer},
high--fidelity quantum gates, and highly parallel manipulation
capabilities~\cite{evered2023high,tsai2025benchmarking,lin2025ai,ma2023high}.
The flexible geometry of optical tweezer arrays, together with tunable
Rydberg interactions, allows programmable connectivity and
interactions to be engineered between atoms~\cite{manovitz2024quantum}.
These features make neutral atom arrays a promising platform for quantum
computing and quantum simulation~\cite{saffman2016quantum,scholl2021quantum}.
They have also been widely used to investigate quantum many-body state
preparation~\cite{ebadi2021quantum}, quantum phase transitions and
nonequilibrium dynamics~\cite{bluvstein2021controlling,semeghini2021probing},
providing a versatile setting for studying and controlling interacting
multi--atom quantum systems.

These systems can all be regarded as point-like atoms with field-regulated interactions. However, this description holds only when the atomic finite spatial scale is negligible; once this scale becomes non-negligible for light-matter coupling, this approximation no longer applies.
When the spatial extent of an artificial atom becomes comparable
to the wavelength of the field to which it couples, the dipole
approximation breaks down. The light--matter interaction must then
account for nonlocal coupling at spatially separated points, leading
to the concept of a giant atom~\cite{guo2017giant,andersson2019non, PhysRevA.109.063708,wang2022giant}. The propagation and phase accumulation between different coupling points
give rise to interference effects that are absent for conventional small
atoms. As a consequence, giant atom systems can exhibit
frequency dependent relaxation rates and Lamb shifts~\cite{frisk2014designing}, non-Markovian dynamics, suppressed
decoherence~\cite{kockum2018decoherence}, bound states in the continuum~\cite{guo2020oscillating}, and chiral photon scattering~\cite{chen2022nonreciprocal}. The internal degree of freedom of a giant atom can, in many situations,
be described as an effective two-level system, for which quantum control
protocols can be designed in a relatively simple and transparent manner.
In realistic settings, however, a controllable quantum subsystem is
generally not completely isolated, but interacts with additional quantum
degrees of freedom. When a giant atom is coupled at multiple sites to a
waveguide composed of interacting atoms, the resulting system becomes a
composite many-atom system involving the controlled giant atom, the
internal dynamics of the atomic waveguide and the coupling between them.
The coupling modifies the eigenstates, energy spectrum, and available
dynamical channels of the full system. Consequently, a control protocol
designed solely for the isolated giant atom does not, in general, remain
an exact control protocol for the coupled system. This raises a natural
question of how fast and high-fidelity quantum control designed for a
simple few-level subsystem can be extended to a more complex interacting
quantum system.

Adiabatic control provides a robust approach to quantum state preparation
and transfer, but a sufficiently slow evolution is generally required to
satisfy the adiabatic condition. Shortcuts to adiabaticity (STA)
\cite{chen2010fast}
have been developed being a family of analytical techniques to design
quantum control schemes which are able
to reproduce the desired outcome of adiabatic evolution within a much
shorter time.
STA techniques have been applied to atom and ion transport,
quantum state preparation, population transfer, quantum gates, quantum
thermodynamics, many-body spin systems and open quantum systems
\cite{STAreview2013,guery2019shortcuts},
including experimental implementations as for example
\cite{zhou2020experimental,yin2022short,hwang2025fast}.

For simple two- or few-level systems, STA protocols can often be
constructed analytically from a prescribed reference
Hamiltonian \cite{chen2010shortcut,Ruschhaupt_2012}. However, when additional interacting degrees of freedom are
included, the Hilbert space and dynamical structure of the system become
considerably more complicated. An STA protocol designed for a small
subsystem may then cease to be exact for the full coupled Hamiltonian,
whereas constructing a new exact STA directly for the entire composite
system can require substantially more complicated control terms.
Enhanced shortcuts to adiabaticity (eSTA) provide an alternative strategy~\cite{whitty2020quantum,whitty2022improved}.
Instead of reconstructing an exact shortcut
for the full system from the beginning, eSTA starts from a known STA
protocol and introduces systematic corrections according to the dynamics
of the complete system, while retaining the basic structure of the
original control scheme.

The formalism of enhanced STA has been applied to atomic transport
in optical lattices \cite{whitty2020quantum,whitty2022robust,hauck2021,hauck2022},
to anharmonic trap expansion \cite{whitty2022improved}
and to spin squeezing in bosonic Josephson junctions \cite{odelli2023,odelli2024}. 

In this work, we investigate fast quantum control of a giant atom coupled
at two spatially separated sites to a one-dimensional interacting atomic
waveguide. We first construct an STA protocol for the isolated giant atom
using Lewis--Riesenfeld invariant-based inverse engineering which plays as
a reference protocol for the full coupled system. We then investigate how
the coupling to the atomic waveguide modifies the performance of the
original STA protocol. To identify the origin of the resulting fidelity
loss, we consider a weak coupling perturbative analysis that separates the leading infidelity into contributions associated with coupling-induced
corrections to the boundary states, dynamical transitions during the
evolution, and the interference between them. This analysis further
relates the control error to the eigenstates and energy gaps of the atomic
waveguide.

To reduce the fidelity loss discussed above, we further employ
the $\mathrm{eSTA}$ scheme~\cite{whitty2022improved}.
We investigate the performance of the resulting control scheme for different waveguide driving strengths,
boundary conditions. Our results show that
the performance of the STA protocol in the composite system is determined
not only by the giant atom waveguide coupling strength but also by the
internal spectral structure of the waveguide. The eSTA correction provides
a systematic way to improve the original shortcut without constructing an
exact STA protocol for the full system.

In Sec.~\ref{model}, we introduce the model and discuss its experimental feasibility.
In Sec.~\ref{resultSTA}, we apply the reference STA protocol to the coupled system and analyze the coupling-induced fidelity loss.
In Sec.~\ref{resultESTA}, we construct the eSTA correction and compare the STA and eSTA fidelities for uncoupled and coupled initial and target states under periodic and open boundary conditions.
In Sec.~\ref{conclusion}, we summarize our main conclusions.

\section{\label{model}Model}

\begin{figure}[t]
    \centering
    \includegraphics[width=1\linewidth]{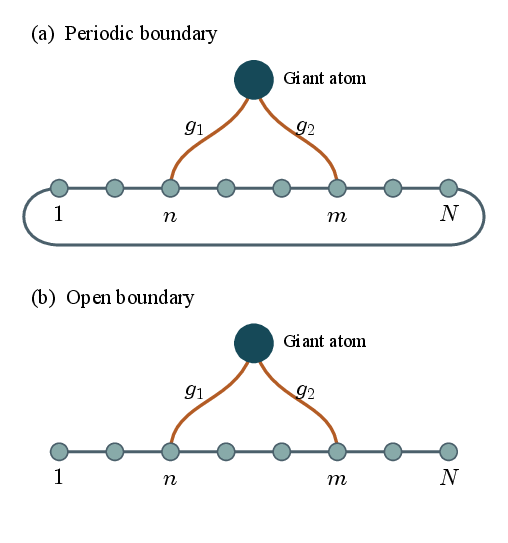}
    \caption{
    Schematic of a giant atom coupled at two spatially separated sites
    to a one-dimensional interacting atomic waveguide with
    (a) periodic and (b) open boundary conditions.
    The giant atom couples to sites $n$ and $m$ with coupling strengths
    $g_1$ and $g_2$, respectively.
    The waveguide contains $N$ two-level atoms.
    For the periodic geometry, the additional link between sites $N$
    and $1$ closes the atomic chain.
    }
    \label{fig:model}
\end{figure}

We consider a two-level giant atom coupled at two spatially separated
sites to a one-dimensional interacting atomic waveguide, as illustrated
in Fig.~\ref{fig:model}. The atomic waveguide consists of $N$ driven
two-level atoms and can be arranged with either open boundary condition
(OBC) or periodic boundary condition (PBC). The giant atom is coupled
to the $n$th and $m$th atoms of the waveguide with coupling strengths
$g_1$ and $g_2$, respectively.

The giant atom is modeled as an effective two-level system with ground
and excited states $|g_G\rangle$ and $|e_G\rangle$, respectively.
It is driven by an external field with a time-dependent Rabi frequency $\Omega(t)$ and detuning $\Delta(t)$.
Within the rotating-wave description, its Hamiltonian is written as
\begin{align}
    H_G(t)
    =
    \frac{\hbar}{2}
    \left[
        \Delta(t)\tau_z
        +
        \Omega(t)\tau_x
    \right],
    \label{HG}
\end{align}
where $\tau_x$ and $\tau_z$ are the Pauli operators of the giant atom, and
$\tau_+=|e_G\rangle\langle g_G|$ and
$\tau_-=|g_G\rangle\langle e_G|$ denote the corresponding raising and lowering operators.

The atomic waveguide consists of $N$ identical driven two-level atoms.
We first consider PBC. For a Rydberg atomic waveguide,
the two internal states of the $i$th atom are denoted by
$|g_i\rangle$ and $|r_i\rangle$. The atoms are driven with Rabi frequency
$\Omega_c$ and detuning $\delta$, and interact with their nearest
neighbors through a van der Waals interaction. The corresponding
Hamiltonian is
\begin{align}
    H_c
    =
    \sum_{i=1}^{N}
    \hbar
    \left(
        \Omega_c \sigma_x^i
        -
        \delta \sigma_{rr}^i
    \right)
    +
    V
    \sum_{i=1}^{N}
    \sigma_{rr}^i\sigma_{rr}^{i+1},
    \label{Hc_vdW}
\end{align}
where
$\sigma_{rr}^i=|r_i\rangle\langle r_i|$,
$\sigma_+^i=|r_i\rangle\langle g_i|$,
$\sigma_-^i=|g_i\rangle\langle r_i|$, and
$\sigma_x^i=\sigma_+^i+\sigma_-^i$.
Here, $V$ denotes the nearest-neighbor van der Waals interaction
strength. For PBC, we identify the $(N+1)$th site with the first one,
i.e., $\sigma_\mu^{N+1}\equiv\sigma_\mu^1$.
For OBC, the interaction between sites
$N$ and $1$ is removed, so that the last sum in
Eq.~\eqref{Hc_vdW} runs from $i=1$ to $N-1$.

The interaction term between giant atom and atomic waveguide is written as
\begin{align}
    H_{\mathrm{int}}
    &=
    g
    \left(
        \sigma_+^n\tau_-
        +
        \sigma_-^n\tau_+
    \right)+
    g
    \left(
        \sigma_+^m\tau_-
        +
        \sigma_-^m\tau_+
    \right),
    \label{Hint}
\end{align}
where $g$ is the coupling strength at the two coupling
sites. $\sigma_n^{\pm}$ and $\sigma_m^{\pm}$ are the raising and lowering operators at the two coupling points of the atomic waveguide. 
Here, we consider identical coupling strength,
$g$ at both sites.  Combining the driven giant atom, the interacting atomic waveguide, and
their two-point coupling, the total Hamiltonian of the system is
\begin{align}
    H(t)
    =
    H_G(t)
    +
    H_c
    +
    H_{\mathrm{int}},
    \label{Htotal}
\end{align}
where $H_c$
describes the
waveguide with van der Waals
interactions.

\subsection{\label{exper}Experimental feasibility}

A possible realization of the present model is a hybrid
superconducting--neutral-atom platform. Superconducting artificial
giant atoms with multiple spatially separated coupling points have been
demonstrated experimentally
~\cite{kannan2020waveguide,andersson2019non,vadiraj2021engineering},
while coherent coupling between ultracold or Rydberg atoms and
superconducting microwave resonators has also been realized
~\cite{hattermann2017coupling,morgan2020coupling}.

For the van der Waals interaction, a possible realization is a
one-dimensional array of neutral ${}^{87}\mathrm{Rb}$ atoms trapped by
optical tweezers. The effective two-level states can, for example, be chosen as
$|g\rangle=|5S_{1/2},F=2,m_F=2\rangle$ and
$|r\rangle=|70S_{1/2},m_J=1/2\rangle$.
The transition between these states can be driven through a two-photon process. When two atoms are excited to the Rydberg state, they interact through a van der Waals potential proportional to $C_6/R^6$, which
provides the interaction term
$V\sum_i\sigma_{rr}^{i}\sigma_{rr}^{i+1}$ used in our model.
Large programmable Rydberg-atom arrays with coherent optical driving
and strong van der Waals interactions have already been realized
experimentally
~\cite{scholl2021quantum,ebadi2021quantum,bluvstein2021controlling}.

The geometry of optical-tweezer arrays can also be programmed over a
wide range of configurations. A linear array provides a natural
realization of open boundary conditions, while a ring-shaped array can be used to realize periodic boundary conditions. Rydberg quantum simulators with one-dimensional ring geometries have recently been demonstrated experimentally
~\cite{fang2025probing}. In the present model, only the nearest-neighbor van der Waals interaction is retained for simplicity.

\section{\label{resultSTA} STA control protocol in the combined system}

\subsection{Reference STA protocol}

\begin{figure*}[htbp]
    \centering
\includegraphics[width=0.9\linewidth]{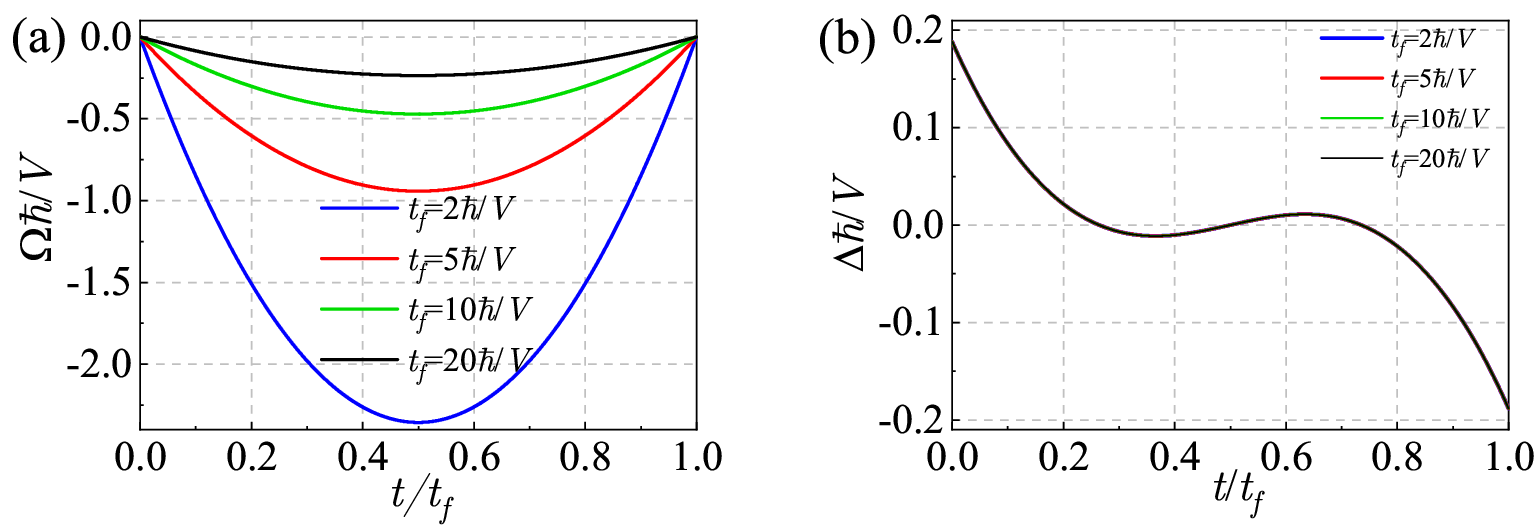}
\caption{
STA control fields designed for the isolated giant atom.
(a) Rabi frequency $\Omega(\tau)$ and
(b) detuning $\Delta(\tau)$ as functions of the normalized time
$\tau=t/t_f$. $t_f=2 \hbar/V$ (blue lines), $t_f=5 \hbar/V$ (red lines),$t_f=10 \hbar/V$ (green lines), $t_f=20 \hbar/V$ (black lines); in (b) all lines are on top of each other.
$\Delta_0=3\pi V/50 \hbar $. 
}
\label{fig:control_function}
\end{figure*}

We first construct an STA protocol for the
isolated giant atom and use it as the reference control throughout this
work. For the two-level giant atom described by Eq.~\eqref{HG}, the STA
protocol is obtained using Lewis--Riesenfeld invariant-based inverse
engineering~\cite{Ruschhaupt_2012}. A dynamical invariant $I(t)$ satisfies
\begin{align}
    \frac{\partial I(t)}{\partial t}
    +\frac{1}{i\hbar}[I(t),H_G(t)]
    =0,
\end{align}
and can be parameterized in terms of two auxiliary functions
$\gamma(t)$ and $\beta(t)$ as
\begin{align}
    I(t)
    =
    \frac{1}{2}
    \begin{pmatrix}
        \cos\gamma(t)
        &
        \sin\gamma(t)e^{i\beta(t)}
        \\
        \sin\gamma(t)e^{-i\beta(t)}
        &
        -\cos\gamma(t)
    \end{pmatrix},
    \label{invariant_result}
\end{align}

The instantaneous eigenstates of the invariant are
\begin{subequations}
\begin{align}
    |\phi_+(t)\rangle
    &=
    \cos\left(\frac{\gamma(t)}{2}\right)
    e^{i\beta(t)}|e_G\rangle
    +
    \sin\left(\frac{\gamma(t)}{2}\right)|g_G\rangle,
    \\
    |\phi_-(t)\rangle
    &=
    \sin\left(\frac{\gamma(t)}{2}\right)|e_G\rangle
    -
    \cos\left(\frac{\gamma(t)}{2}\right)
    e^{-i\beta(t)}|g_G\rangle ,
\end{align}
\end{subequations}
In general,
$|\phi_\pm(t)\rangle$ themselves do not satisfy the time-dependent
Schr\"odinger equation. The corresponding exact dynamical solutions
are obtained by including the Lewis--Riesenfeld phases,
$e^{i\alpha_\pm(t)}|\phi_\pm(t)\rangle$, where
\begin{equation}
\alpha_\pm(t)=\frac1\hbar\int_0^t
\langle\phi_\pm(t')|
i\hbar\partial_{t'}-H_G(t')
|\phi_\pm(t')\rangle\,\mathrm dt' ,
\label{eq:LR_phase}
\end{equation}
In the following, we choose
$e^{i\alpha_+(t)}|\phi_+(t)\rangle$
as the reference STA evolution of the giant atom.

The auxiliary functions are related to the physical control fields by
\begin{subequations}
\begin{align}
    \Omega(t)
    &=
    \frac{\dot{\gamma}(t)}
    {\sin\beta(t)},
    \\
    \Delta(t)
    &=
    \frac{\dot{\gamma}(t)
    \cos\gamma(t)\cos\beta(t)}
    {\sin\gamma(t)\sin\beta(t)}
    -
    \dot{\beta}(t) ,
\end{align}
 \label{STA_control_relation}
\end{subequations}

For the population-inversion process considered here, the giant atom
starts from $|g_G\rangle$ and is transferred to $|e_G\rangle$.
Accordingly, the auxiliary parameters are chosen such that
$\gamma(0)=\pi$ and $\gamma(t_f)=0$. We also require $0<\beta(t)<\pi$. 
In addition, we require the following endpoint conditions of
the physical control fields
\begin{align}
    \Omega(0)=\Omega(t_f)=0,
    \qquad
    \Delta(0)=\Delta_0,
    \qquad
    \Delta(t_f)=-\Delta_0 ,
    \label{STA_boundary_result}
\end{align}
which leads to the additional conditions
$\dot\gamma(0)=\dot\gamma(t_f)=0$, $\beta(0)=\beta(t_f)=\pi/2$, $\dot\beta(0)=-\Delta_0/3$, $\dot\beta(t_f)=\Delta_0/3$.
In addition, we also set $\beta(t_f/2)=\pi/2$.
A convenient choice satisfying these conditions is the following polynomials:
\begin{subequations}
\begin{align}
    \gamma(t)
    &=
    \pi
    \left(
        1-3\tau^2+2\tau^3
    \right),
    \\
    \beta(t)
    &=
    \frac{\pi}{2}
    +
    \frac{\Delta_0 t_f}{3}
    \left(
        -\tau+5\tau^2-8\tau^3+4\tau^4
    \right),
\end{align}
\label{STA_auxiliary_functions}
\end{subequations}
where $\tau=t/t_f$ denotes the normalized time.
Substitution of Eq.~\eqref{STA_auxiliary_functions} into
Eq.~\eqref{STA_control_relation} gives the STA control fields applied
to the giant atom.

Fig.~\ref{fig:control_function} shows the STA control fields as functions of the normalized time $\tau=t/t_f$ for different total evolution times. As shown in Fig.~\ref{fig:control_function}(a), the Rabi frequency $\Omega(\tau)$ vanishes at both endpoints and remains negative throughout the intermediate evolution, reaching its minimum at $\tau=1/2$. Its absolute amplitude increases as $t_f$ decreases, indicating that a shorter protocol requires stronger driving.

The detuning $\Delta(\tau)$, shown in Fig.~\ref{fig:control_function}(b), connects the fixed endpoint values $\Delta_0$ and $-\Delta_0$, where $\Delta_0=3\pi V/(50\hbar)$. The detuning varies nonmonotonically, initially decreasing below zero and then rising through zero near the midpoint to a small positive maximum before decreasing again to its final negative value. Unlike the Rabi frequency, the detuning exhibits nearly identical profiles for all evolution times considered, with the curves overlapping at the scale of the figure. These control fields constitute the reference STA protocol designed for the isolated two-level giant atom.


\subsection{Reference STA protocol in the coupled system}

The above STA protocol is constructed  from the Hamiltonian of
the isolated giant atom. We now ask how this protocol performs when the
giant atom is embedded in the interacting atomic waveguide introduced
in Sec.~\ref{model}.

In detail, we will consider two settings. 
In the first setting we assume that the giant atom and the waveguide are only coupled during the evolution for $0<t<t_f$. The motivation is that the coupling to the waveguide models an (un-wanted) coupling to a structured environment. Therefore, the system is assumed to be in the ground state $|\Psi_{0,g=0}\rangle$ corresponding to the uncoupled system at initial time $t=0$ (which is a product state). We also want the system to end up in the corresponding ground state of the uncoupled system at final time $t=t_f$, the target state $|\Psi_{T,g=0}\rangle$.
The performance of the reference protocol is characterized by the
final-state fidelity
\begin{align}
    F_0
    =
    \left|
        \langle \Psi_{T,g=0}|\Psi(t_f)\rangle
    \right|^2,
    \label{eq_fidelity0}
\end{align}
where $|\Psi(t_f)\rangle$ is obtained by evolving the initial ground
state $|\Psi_{0,g=0}\rangle$ of the uncoupled Hamiltonian at $t=0$, and
$|\Psi_{T,g=0}\rangle$ is the ground state of the uncoupled Hamiltonian at the
final time $t_f$. 

In the second setting, we are considering a state transfer from the ground state of the coupled system with $\Delta(0)=\Delta_0$ to the ground state of the coupled system with $\Delta(t_f)=-\Delta_0$. Here, the giant atom and the waveguide are coupled all the time. 
The performance of the reference protocol is characterized by the
final-state fidelity
\begin{align}
    F
    =
    \left|
        \langle \Psi_{T,g>0}|\Psi(t_f)\rangle
    \right|^2,
    \label{full_fidelity}
\end{align}
where $|\Psi(t_f)\rangle$ is obtained by evolving the initial ground
state $|\Psi_{0,g>0}\rangle$ of the full, coupled Hamiltonian at $t=0$, and
$|\Psi_{T,g>0}\rangle$ is the ground state of the full, coupled Hamiltonian at the
final time $t_f$. 

\begin{figure}[t]
\centering

\includegraphics[width=0.8\linewidth]{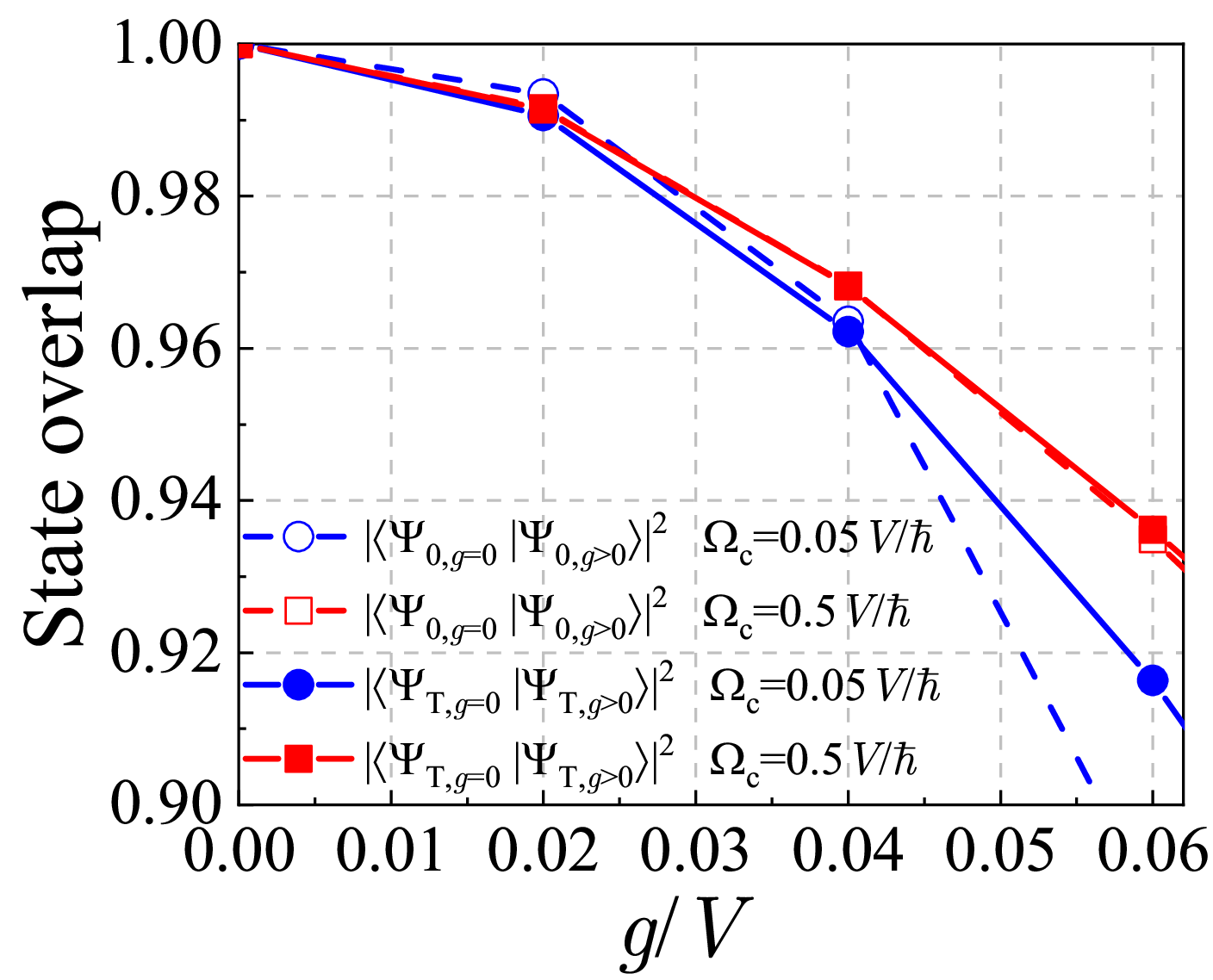}
\caption{Overlaps between the coupled and uncoupled ground states at the initial and final times as functions of the coupling strength $g/V$, for $\Omega_c=0.05V/\hbar$ and $0.5V/\hbar$. The atomic waveguide is subject to periodic boundary conditions.
\label{fig:overlap}
}
\end{figure}

These two settings are obviously different, this can be also seen in Fig.~\ref{fig:overlap} which shows the overlap of the two initial states,
$ |\langle\Psi_{0,g=0}|\Psi_{0,g>0}\rangle|^2 $ (dashed lines)
as well as the overlap of the two target states 
$ |\langle\Psi_{T,g=0}|\Psi_{T,g>0}\rangle|^2 $ (solid lines). As expected, the overlap decreases with increasing coupling between the giant atom and the wave guide.

In both settings, first the same functions $\Omega(t)$ and $\Delta(t)$ for the giant atom derived in the previous subsection and shown in
Fig.~\ref{fig:control_function}. The
time evolution of the actual state evolves according to the full coupled Hamiltonian
Eq.~\eqref{Htotal}.
We are considering here the giant atom coupled
to a van der Waals atomic waveguide under periodic boundary
conditions.

\begin{figure*}[t]
\centering

\includegraphics[width=0.9\linewidth]{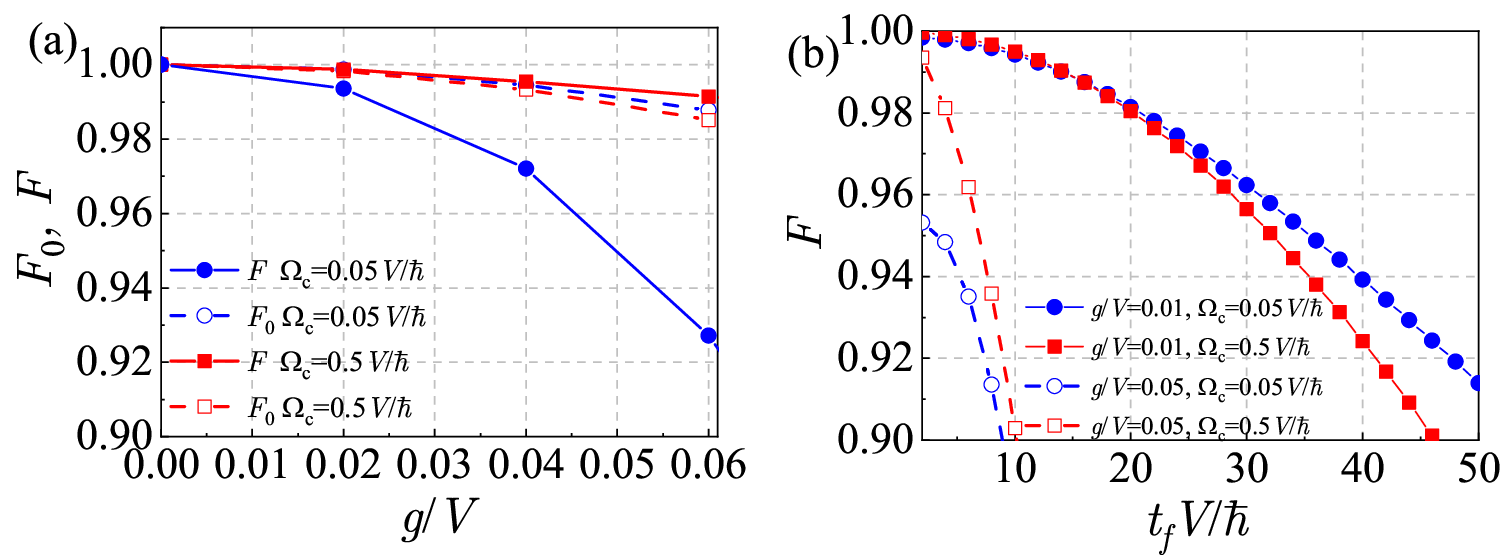}

\caption{Reference STA protocol for the giant atom coupled
to atomic waveguide under periodic boundary
conditions.
(a) $F_0$ and $F$ as functions of the coupling strength $g/V$ at $t_f=2\hbar/V$, for $\Omega_c=0.05V/\hbar$ and $0.5V/\hbar$. The initial and target states are the uncoupled ground states for $F_0$ and the coupled ground states for $F$, both fidelities are evaluated using the full coupled dynamics. (b) $F$ as a function of the evolution time $t_fV/\hbar$ for $g/V=0.01$ and $0.05$ with $\Omega_c=0.05V/\hbar$ and $\Omega_c=0.5V/\hbar$.
\label{fig:STAalone}
}
\end{figure*}

In Fig.~\ref{fig:STAalone}(a), the total evolution time is fixed at
$t_f=2\hbar/V$ and the coupling strength is varied for two different
waveguide Rabi frequencies. The results of both settings, characterized by the fidelities $F_0$ and $F$ are shown.
Note that for $g=0$ the fidelities $F_0$ and $F$ must be always one which follows also from the construction of the control functions.

As expected, both fidelities decrease for increasing $g$.
In general, we see here that the influence of the waveguide cannot be characterized by the coupling strength alone, but also by the Rabi frequency of the waveguide $\Omega_c$ plays a role. The fidelity $F_0$ depends only weakly on $\Omega_c$ and does not decrease too drastically with increasing $g$.

On the other hand,  fidelity $F$ shows a much stronger dependence on $\Omega_c$. For $\Omega_c=0.5V/\hbar$, the fidelity $F$ decreased also not too drastically with increasing $g$.
On the other hand, for $\Omega_c=0.05V/\hbar$, the fidelity $F$ decreases noticeably more as $g$
increases, reaching a substantially lower value for $g/V=0.06 $ than the in the case with $\Omega_c=0.5V/\hbar$.
This shows that the sensitivity of the reference STA protocol to the giant atom waveguide coupling strongly depends on the internal
parameters of the atomic waveguide, especially for the fidelity $F$.

We are examining this fidelity $F$ further in Fig.~\ref{fig:STAalone}(b); this figure shows the fidelity $F$ as a function of the total evolution time for different giant atom waveguide coupling strengths.
Again, when $g=0$, the giant atom and the atomic waveguide are decoupled; the waveguide therefore does not interfere with the STA process, and
the $F$ is one for all $t_f$ by construction and the desired state transfer of the isolated giant atom is realized.

Once the coupling is introduced, the same STA control fields are no
longer exact for the full composite system. In general, for
$g=0.01V$, the deviation of $F$ from unity remains smaller over the time interval shown, whereas a substantially larger reduction is
observed for $g=0.05V$.
Fidelity $F$ shows again the dependence on the Rabi frequency $\Omega_c$.
It is interesting to see that the the fidelity is also decreasing with increasing final time $t_f$ in the shown range of final times.
This can be explained that the perturbation by the waveguide has more time to disturb the evolution of the giant atom.

The results in Fig.~\ref{fig:STAalone} reveal two distinct aspects of the
control problem. First, the STA protocol constructed for the isolated
giant atom provides an exact reference in the absence of coupling but
generally ceases to be exact once the giant atom is coupled to the
interacting waveguide. Second, the resulting fidelity loss is not
determined solely by the coupling strength $g$, but also by the internal parameters of the waveguide. This raises the question of which
coupling-induced processes are responsible for the loss of fidelity
and why their contributions depend so strongly on $\Omega_c$.
To answer this question, we next analyze the fidelity loss in the
weak-coupling regime using perturbation theory.


\subsection{Approximation of the coupling-induced fidelity loss}

The results in Fig.~\ref{fig:STAalone} show that the fidelity loss
depends not only on the coupling strength $g$,
but also strongly on the waveguide Rabi frequency $\Omega_c$.
To understand the origin of this dependence, we analyze the
full-system dynamics perturbatively in the weak-coupling regime.
We write the total Hamiltonian as
\begin{align}
    H(t)=H_0(t)+gW_{\mathrm{int}},
\end{align}
where
$H_0(t)=H_G(t)+H_c$
describes the uncoupled giant atom and atomic waveguide, while
$gW_{\mathrm{int}}$ is treated as the perturbation.

For weak coupling, the final fidelity can be approximated as (see Appendix for details)
\begin{eqnarray}
    F &=& 1 - g^2 \cR + O(g^3),
\end{eqnarray}
Note that $g$ has a unit of energy here and $\cR$ has a unit of 1/energy$^2$.
We have $\cR=\cR_d + \cR_i + \cR_T + \cR_0$
and
\begin{eqnarray}
\cR_d&=& -\fabsq{\braket{\Phi_T^{(0)}}{U^{(1)}\Phi_0^{(0)}}} - 2 \mbox{Re} \braket{\Phi_T^{(0)}}{U^{(2)}\Phi_0^{(0)}},\\
\cR_0 &=& 
-\fabsq{\braket{\Phi_0^{(0)}}{\Phi_0^{(1)}}} + \fnormq{\Phi_0^{(1)}}\nonumber\\
&\approx& \frac{1}{g^2}\left(1-\fabsq{\braket{\Phi_{0}^{(0)}} {\Phi_{0}(g)}}\right),\\
\cR_T &=& 
-\fabsq{\braket{\Phi_T^{(0)}}{\Phi_T^{(1)}}} + \fnormq{\Phi_T^{(1)}} \nonumber\\
&\approx& \frac{1}{g^2} \left(1-\fabsq{\braket{\Phi_{T}^{(0)}} {\Phi_{T}(g)}}\right),
\end{eqnarray}
and
\begin{eqnarray}
\cR_{i}&=& -2 \mbox{Re}\Big(
\braket{\Phi_0^{(0)}}{\Phi_0^{(1)}}\braket{U^{(1)}\Phi_0^{(0)}}{\Phi_T^{(0)}}\nonumber\\
& & + \braket{\Phi_0^{(0)}}{\Phi_0^{(1)}}\braket{\Phi_T^{(0)}}{\Phi_T^{(1)}}\nonumber\\
& & + \braket{\Phi_T^{(1)}}{\Phi_T^{(0)}}\braket{U^{(1)}\Phi_0^{(0)}}{\Phi_T^{(0)}}\nonumber\\
& & + \braket{\Phi_T^{(1)}}{U^{(0)}\Phi_0^{(1)}}
+ \braket{\Phi_T^{(1)}}{U^{(1)}\Phi_0^{(0)}}\nonumber\\
& & + \braket{\Phi_T^{(0)}}{U^{(1)}\Phi_0^{(1)}}
\Big) ,
\end{eqnarray}
The detailed derivation is given in Appendix~\ref{perturbation}.

Here $\cR_0$ and $\cR_T$
are the contributions from the
coupling-induced corrections to the initial and target states,
$\mathcal{R}_{d}$ describes the transitions accumulated
during the dynamical evolution, and $\mathcal{R}_{\mathrm{i}}$ represents
the interference between these two contributions.

For further evaluation of these terms, we will choose a convenient basis, we expand the perturbative corrections in the eigenbasis
of the uncoupled system. 

The waveguide eigenstates and eigenvalues
satisfy
\begin{align}
    H_c|\psi_l\rangle=E_l|\psi_l\rangle,
\end{align}
with $E_l\leq E_{l+1}$.
The construction of the correction uses the following reference
solutions of the uncoupled system
\begin{align}
    |\chi_{l,s}(t)\rangle
    =
    e^{-iE_lt/\hbar}|\psi_l\rangle
    \otimes
    e^{i\alpha_s(t)}|\phi_s(t)\rangle,
\end{align}
where $l=0,1,...$ and $s=\pm$.
The reference channel is $(0,+)$, with product endpoint states
\begin{align}
    \ket{\Psi_{0,g=0}} = \ket{\Phi_0^{(0)}}
    &=
    |\chi_{0,+}(0)\rangle,
    \\
    \ket{\Psi_{T,g=0}}=\ket{\Phi_T^{(0)}}
    &=
    |\chi_{0,+}(t_f)\rangle ,
\end{align}
So the reference evolution corresponds to the channel $(0,+)$, consisting
of the waveguide ground state $|\psi_0\rangle$ and the STA branch
$|\phi_+(t)\rangle$ of the giant atom.

It has been shown \cite{whitty2020quantum} that
\begin{eqnarray}
   \cR_{d}
    &=&
    \frac{1}{\hbar^2}
    \sum_{l,s}^{}|G_{l,s}|^2,
\end{eqnarray}
where $(l,s) \neq (0,+)$ in the sum and
\begin{align}
    G_{l,s}
    &=
    \int_0^{t_f}dt\,
    \langle\chi_{l,s}(t)|
    W_{\mathrm{int}}
    |\chi_{0,+}(t)\rangle ,
    \label{esta_G_reference}
\end{align}

For the interaction considered,
$W_{\mathrm{int}}=S_+\tau_-+S_-\tau_+$,
the transition amplitude can be written as
\begin{align}
    G_{l,s}
    =
    B_l^+A_{s,-}^{l}
    +
    B_l^-A_{s,+}^{l},
    \label{transition_amplitude}
\end{align}
Here, 
the coupling of the reference channel to a waveguide eigenstate
$|\psi_l\rangle$ is characterized by
\begin{align}
    B_l^{\pm}
    &=
    \langle\psi_l|S_\pm|\psi_0\rangle,
    \\
    S_\pm &=\sigma_\pm^n+\sigma_\pm^m ,
    \label{waveguide_matrix_element}
\end{align}
The corresponding time-dependent transition amplitude of the giant atom
is
\begin{align}
    A_{s,\pm}^{l}
    &=
    \int_0^{t_f}dt\,
    \langle\phi_s(t)|\tau^\pm|\phi_+(t)\rangle
    \nonumber\\
    &\quad\times
    \exp\left\{
        i[\alpha_+(t)-\alpha_s(t)]
        +
        \frac{i}{\hbar}(E_l-E_0)t
    \right\},
    \label{giant_atom_amplitude}
\end{align}
where $s=\pm$.

\subsection{Comparison of approximation and numerical results}

We now want to examine the approximation derived in the previous subsection and compare it also with the exact numerical results.

\begin{figure*}[t]
    \centering
\includegraphics[width=0.62\linewidth]{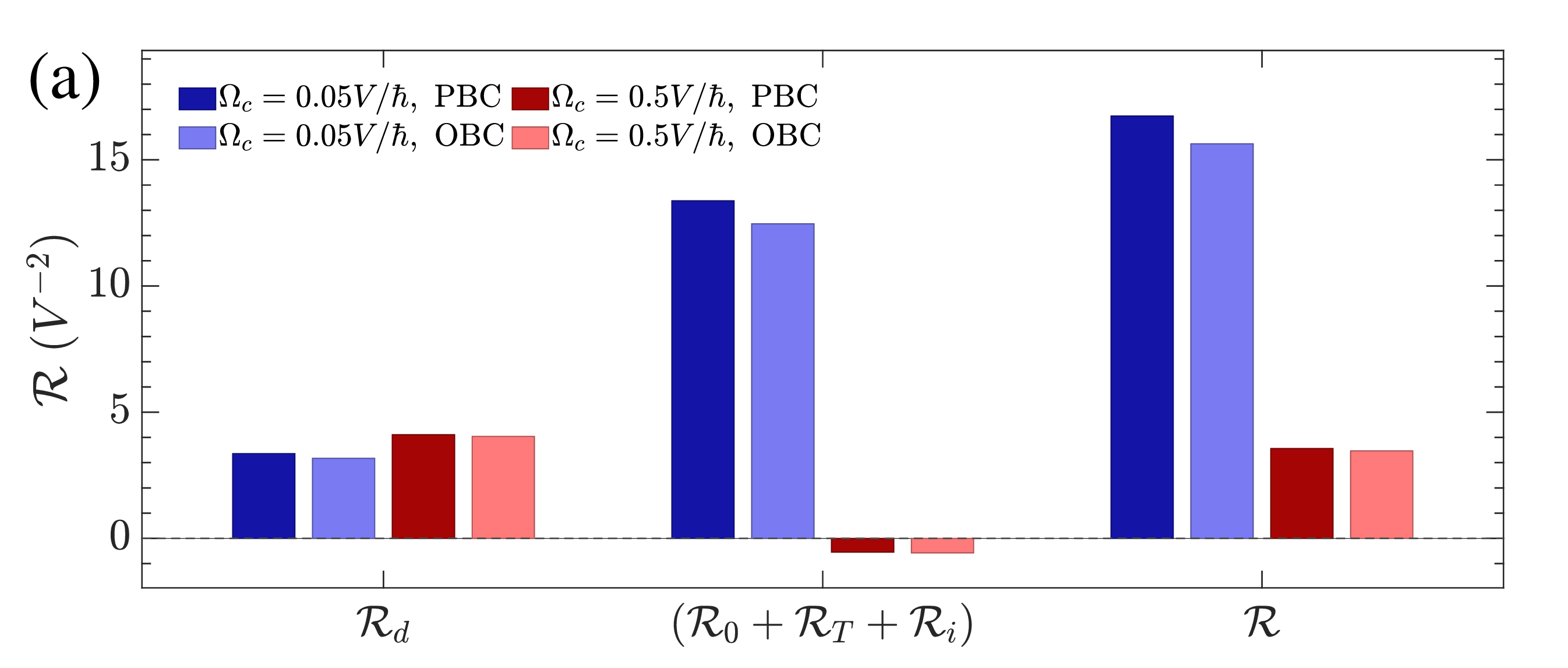}\\
\includegraphics[width=0.82\linewidth]{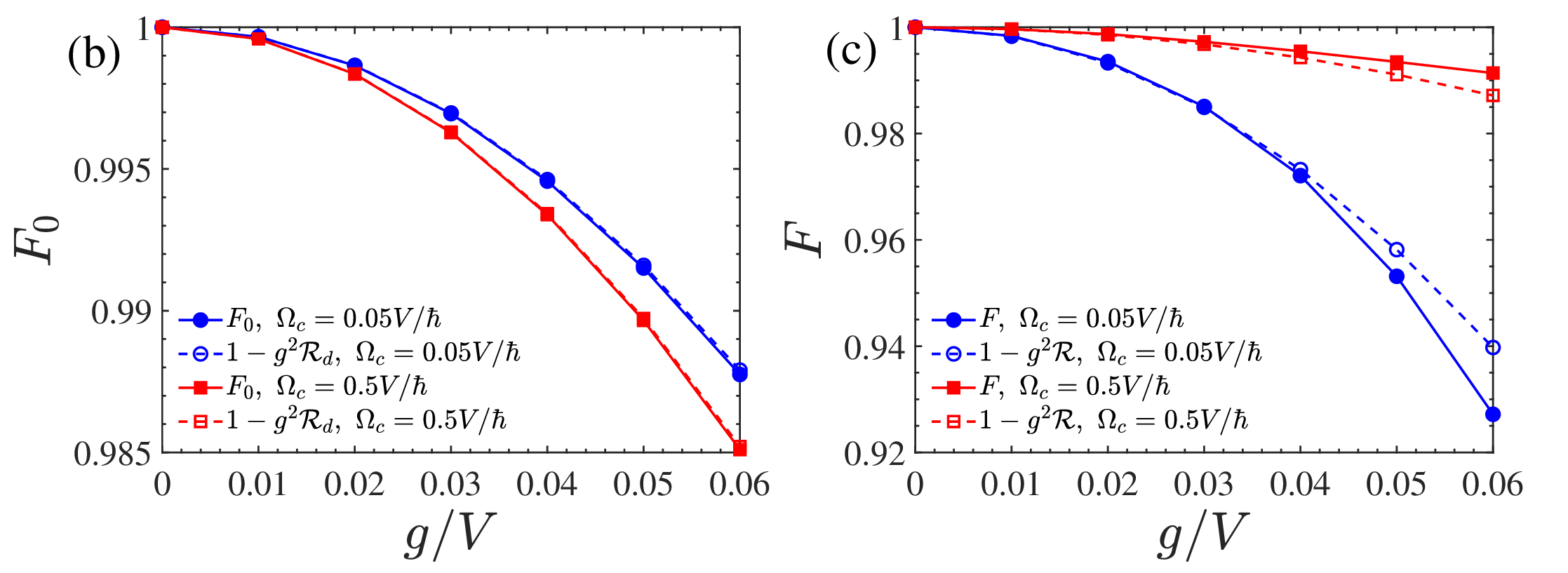}
    \caption{
    Perturbative quantities and leading-order fidelity loss contributions
    for $\Omega_c=0.5V/\hbar$ and $0.05V/\hbar$ at
    $t_f=2\hbar/V$.
    (a) Comparison of the dynamical contribution $\mathcal{R}_d$, the combined contribution $\mathcal{R}_0+\mathcal{R}_T+\mathcal{R}_i$, and the total coefficient $\mathcal{R}$ under periodic and open boundary conditions. (b) Fidelity $F_0$ and its perturbative approximation $1-g^2\mathcal{R}_d$ with different $g/V$. (c) Fidelity $F$ and its perturbative approximation $1-g^2\mathcal{R}$ as functions of $g/V$. Periodic boundary conditions are consider in (b) and (c).
    }
    \label{fig:approx}
\end{figure*}

We first look at the different coefficients contributing to $\cR$
in Fig.~\ref{fig:approx}(a), namely $\cR_d$ and
$\cR_0+\cR_T+\cR_i$.
For calculating these coefficients, we take all states
$\ket{\chi_{l,s}}$ of the system into account.
We compare four different parameter sets: periodic and open
boundary conditions as well as $\Omega_c=0.05V/\hbar, 0.5V/\hbar$.
In general, changing $\Omega_c$ or the boundary conditions modifies
the eigenstates and energy gaps of the atomic waveguide and
therefore also changes these coefficients.
In Fig.~\ref{fig:approx}(a), we see that there is only a little
difference between the cases PBC and OBC for the same value
of $\Omega_c$.

For $\Omega_c=0.5V/\hbar$, the dynamical contribution
$\mathcal{R}_d$ is the largest positive contribution while the sum
$\cR_0 + \cR_T + \cR_i \approx 0$ such that we get that the total coefficient $\cR \approx \cR_d$. This means that the differences between the ideal initial/final states and the corresponding states with $g>0$ is not very significant in this case.

However, for $\Omega_c=0.05V/\hbar$, $\cR_0 + \cR_T + \cR_i$
becomes the dominant contribution and we get that $\cR$ is significant larger than $\cR_d$, and the main loss in fidelity originates from the difference between the ideal initial/final states and the corresponding states with $g>0$ in this case.

As the results for the different boundary conditions for a fixed $\Omega_c$ are not very different, we will now consider only the case of periodic boundary conditions in 
Fig.~\ref{fig:approx}(b) and (c).

In Fig.~\ref{fig:approx}(b), we want now to compare the numerically exact fidelity $F_0$ with the approximation $1-g^2 \cR_d$. Note that in this setting, we start with the ideal initial state and the fidelity is calculated using the ideal target state such that $\cR_0=\cR_T=\cR_i=0$ in this case and only $\cR_d$ is relevant here.
We see a very good agreement between the exact numerical fidelity and the approximation for both values of $\Omega_c$ in this setting
for the shown values of $g$.

In Fig.~\ref{fig:approx}(c), we want now to compare the numerically exact fidelity $F$ in the second setting with the approximation $1-g^2 \cR$. We see that there is a good agreement between the numerical fidelity and the approximation for $g/V < 0.045$.
There is some small deviation for larger values of $g$ showing that some higher-order (in $g$) effects will start to play a role.
The lower fidelity for $\Omega_c=0.05V/\hbar$ results
from the larger difference between the ideal uncoupled
states and the coupled states in this case, as explained above, rather than an increase
in the contribution of dynamical transitions $\cR_d$.

\section{Enhanced STA protocol in the combined system\label{resultESTA}}

The previous results show that the STA protocol designed for the
isolated giant atom is no longer exact when the coupling to the
atomic waveguide is included. To increase the resulting fidelity,
we employ now enhanced Shortcuts to Adiabaticity
(eSTA)~\cite{whitty2020quantum, whitty2022improved}.

\subsection{Review of general eSTA formalism}

Here we will summarize the enhanced STA formalism taking advantage of Hessian, which was introduced in
Ref.~\cite{whitty2022improved} while using the notation from \cite{odelli2023,odelli2024}.

Let us assume a system is described by a Hamiltonian $H_{\mu} (\blambda,t)$ where $\mu$ is a system parameter and $\blambda$ some parameterization of a control function. We assume that for $\mu\to 0$, i.e., for $H_0 (\blambda,t)$, the STA framework can be applied to the system and an STA control function, parameterized as $\blambda_{\mathrm{STA}}$, can be constructed, as well as the STA solutions' evolution in time is known. 
We label STA solutions $\{\ket{\chi_n (t)}\}$, with $\ket{\chi_n (t)} = U_{0,\bnull} (t,0) \ket{\chi_n (0)}$.

However, if $\mu \neq 0$, then the STA control scheme does not give perfect fidelity anymore. In the formalism of eSTA, a correction $\bepsilon$ to the STA control scheme $\vec\lambda_{\mathrm{STA}}$ can be analytically constructed~\cite{whitty2020quantum, whitty2022improved}, such that $\vec\lambda=\vec\lambda_{\mathrm{STA}} + \vec\epsilon$ improves the fidelity with $\mu \neq 0$. In the following, we will just review the final formula.
To simplify the notation, we assume $\blambda_{\mathrm{STA}} = 0$ and so $\blambda=\bepsilon$.

We assume that the initial and final ground states of $H_\mu$ can be approximated by the 
ground states of $H_0$.
Therefore, the performance of the corrected controls in the full coupled
system is evaluated using the fidelity
\begin{align}
    F(\mu,\blambda)
    =
   | \braket{\chi_0(t_f)}{
        U_{\mu,\blambda}(t_f,0)|
        \chi_0}|^2,
    \label{esta_fidelity}
\end{align}
where we assume that we are interested in a state transfer along $\chi_0$ in the approximated system and
$U_{\mu,\blambda}(t_f,0)$ is generated by the Hamiltonian $H_\mu(\blambda, t)$.

Now we define
\begin{align}
\Gamma_{n,m}(t) = \braXket{\chi_n(t)}{\Delta H_{\mu}(\blambda,t)}{\chi_m(t)},
\end{align}
where
$\Delta H_{\mu}(\blambda,t) = H_{\mu} (\blambda,t) - H_0 (\bnull,t)$

Let
\begin{eqnarray}\label{eq:Gn}
\cG_n &=& \frac{1}{\hbar} \int_0^{t_f} dt \: \Gamma_{n,0}(t)\bigg|_{\mu,\blambda=\bnull}\nonumber\\
&=& \frac{1}{\hbar} \int_0^{t_f} dt \:
\braXket{\chi_n(t)}{\Delta H_{\mu}(\bnull,t)}{\chi_0(t)},
\end{eqnarray}
and a vector $\bK$, where
the $k^{\text{th}}$ component of $\bK_n$ is given by
\begin{eqnarray}\label{eq:Kn}
\left(\bK_n\right)_k &=& \frac{1}{\hbar} \int_0^{t_f} dt \: \frac{\partial}{\partial \lambda_k}\Gamma_{n,0} (t)\bigg|_{\mu,\blambda=\bnull}\nonumber\\
&=& \frac{1}{\hbar} \int_0^{t_f} dt \: 
\braXket{\chi_n(t)}{\frac{\partial H_{\mu}}{\partial \lambda_k} (\bnull,t)}{\chi_0(t)},
\end{eqnarray}
Finally, let
\begin{align}\label{eq:Hess_approx}
\mathbb{H}_{l,k} =
& \sum_{n=1}^{N_{cut}}
\text{Re} \left[ \cG_n^* \:
\left(W_n\right)_{l,k}
+
\left( \bK_n^* \right)_l \: \left( \bK_n \right)_k
\right],
\end{align}
with
\begin{align}\label{eq:Wn}
\left(W_n\right)_{l,k} = 
\frac{1}{\hbar} \int_0^{t_f} dt \:
\frac{\partial^2}{\partial \lambda_l \partial \lambda_k}\Gamma_{n,0} (t)\bigg|_{\mu,\vec{\lambda}=\vec{0}},
\end{align}

It is important to underline that all these quantities can be calculated for the known STA solutions of the idealised system with $\mu=0$.

It can be shown \cite{whitty2022improved}, that we can write the fidelity and gradient approximations now as
\begin{align}\label{eq:eSTA_fid_Gn}
F \left( \mu, \vec{0} \right) 
\approx 1 - 
\sum_{n=1}^{N_{\mathrm{cut}}} \fabsq{\cG_n}=:f,
\end{align}
the gradient approximation is given by
\begin{align}\label{eq:eSTA_grad_Kn}
\nabla_{\vec{\lambda}} F \left( \mu, \vec{0} \right) 
\approx (-2) \vec{v},
\end{align}
and
\begin{eqnarray}
\vec v &=& \sum_{n=1}^{N_{\mathrm{cut}}} \mbox{Re} \left(\cG_n^* \vec{K}_n\right),
\end{eqnarray}
and we have truncated the infinite sums to the first ${N_{\mathrm{cut}}}$ terms.
In addition, the Hessian of the fidelity can be approximated by
\begin{eqnarray}
    \frac{\partial^2 F}{\partial \lambda_l \partial \lambda_k} \approx (-2)\mathbb{H}_{l,k} ,
\end{eqnarray}

Assuming a parabolic series expansion of the fidelity, the eSTA correction can be calculated as
\begin{align}\label{eq:eSTA_Hessian_final}
\vec{\lambda}_{\mathrm{eSTA}} =
-\frac{\vec{v} \fnorm{\vec{v}}^2}
{\vec{v}^{\:T} \:\mathbb{H} \:\vec{v}},
\end{align}
Further details and derivations can be found in \cite{whitty2020quantum, whitty2022improved}.

\subsection{Application to the giant-atom waveguide system}

We apply the general eSTA formalism to the coupled giant-atom
waveguide system. 
 The coupling strength $g$ plays the role of the
system parameter $\mu$, while the uncoupled system with $g=0$
provides the known STA reference solutions. To reduce the
coupling-induced control influence, we introduce corrections to the
giant-atom detuning and Rabi frequency,
\begin{subequations}
\label{esta_controls}
\begin{align}
    \Delta(\boldsymbol{\lambda},t)
    &=
    \Delta_{\mathrm{STA}}(t)
    +
    \sum_{k=1}^{M}\lambda_k\mathcal{L}_k(t),
    \\
    \Omega(\boldsymbol{\lambda},t)
    &=
    \Omega_{\mathrm{STA}}(t)
    +
    \sum_{k=1}^{M}\lambda_{M+k}\mathcal{L}_k(t),
\end{align}
\end{subequations}
where $\Delta_{\mathrm{STA}}(t)$ and $\Omega_{\mathrm{STA}}(t)$
are the STA control fields for the uncoupled giant atom, and $M$
is the number of correction parameters for each field. The real
control correction vector is
\begin{align}
    \boldsymbol{\lambda}
    =
    (\lambda_1,\ldots,\lambda_M,
    \lambda_{M+1},\ldots,\lambda_{2M})^T,
\end{align}
the first $M$ components modify the detuning and remaining
$M$ components modify the Rabi frequency. The original STA
protocol corresponds to $\boldsymbol{\lambda}=\boldsymbol{0}$.

The dimensionless interpolation functions are defined as
\begin{align}
    \mathcal{L}_k(t)
    &=
    \prod_{\substack{j=0\\j\neq k}}^{M+1}
    \frac{t-t_j}{t_k-t_j},
    \qquad k=1,\ldots,M,
    \\
    t_j&=\frac{jt_f}{M+1},
    \qquad j=0,\ldots,M+1,
\end{align}
where $t_f$ is the total evolution time and $t_j$ are the
interpolation nodes. These functions satisfy
\begin{align}
    \mathcal{L}_k(0)
    =
    \mathcal{L}_k(t_f)
    =0,
    \qquad
    \mathcal{L}_k(t_j)=\delta_{kj},
\end{align}
Thus, the corrections
preserve the initial and final values of the control fields.
The parameters $\lambda_k$ and $\lambda_{M+k}$ specify the
detuning and Rabi-frequency corrections at the node $t_k$,
respectively, and both have units of frequency.

The full Hamiltonian with the corrected controls is
\begin{align}
    H_g(\boldsymbol{\lambda},t)
    =
    H_c
    +
    \frac{\hbar}{2}
    \left[
        \Delta(\boldsymbol{\lambda},t)\tau_z
        +
        \Omega(\boldsymbol{\lambda},t)\tau_x
    \right]
    +
    gW_{\mathrm{int}},
    \label{esta_Hlambda}
\end{align}
and the corresponding reference Hamiltonian is
\begin{align}
    H_{g=0}(\boldsymbol{\lambda}=\boldsymbol{0},t)
    =
    H_c
    +
    \frac{\hbar}{2}
    \left[
        \Delta_{\mathrm{STA}}(t)\tau_z
        +
        \Omega_{\mathrm{STA}}(t)\tau_x
    \right],
\end{align}
The corresponding STA solutions are
\begin{align}
    |\chi_{l,s}(t)\rangle
    =
    e^{-iE_lt/\hbar}|\psi_l\rangle
    \otimes
    e^{i\alpha_s(t)}|\phi_s(t)\rangle,
    \label{esta_reference_solutions}
\end{align}
where the waveguide eigenstates satisfy
\begin{align}
    H_c|\psi_l\rangle=E_l|\psi_l\rangle,
\end{align}
with $l=0,1,\ldots$ labeling the waveguide eigenstates and $l=0$
denoting the waveguide ground state. The index $s=+,-$ labels
the two eigenstates of  Lewis--Riesenfeld
invariant, $\alpha_s(t)$ is the corresponding Lewis--Riesenfeld
phase.

With the branch convention adopted here, the reference evolution
follows $|\chi_{l=0,+}(t)\rangle$. Thus, the reference state
$|\chi_0(t)\rangle$ in the general formalism corresponds to
$|\chi_{l=0,+}(t)\rangle$ in our model, while the remaining
channels are labeled by the composite indices
$(l,s)\neq(0,+)$. 

The full uncoupled reference propagator is constructed from
these solutions as
\begin{align}
    U_{g=0,\boldsymbol{\lambda}=\boldsymbol{0}}(t,t')
    =
    \sum_{l,s}
    |\chi_{l,s}(t)\rangle
    \langle\chi_{l,s}(t')|,
\end{align}
where the sum includes all waveguide eigenstates and both
giant-atom branches.

Following the general definition in the preceding subsection,
the difference between the full and reference Hamiltonians is
\begin{align}
    \Delta H_g(\boldsymbol{\lambda},t)
    &=
    H_g(\boldsymbol{\lambda},t)
    -
    H_{g=0}(\boldsymbol{\lambda}=\boldsymbol{0},t)
    \nonumber\\
    &=
    gW_{\mathrm{int}}
    +
    \frac{\hbar}{2}
    \sum_{k=1}^{M}\mathcal{L}_k(t)
    \left(
        \lambda_k\tau_z+\lambda_{M+k}\tau_x
    \right),
\end{align}
so that, at fixed coupling strength $g$ and
$\boldsymbol{\lambda}=\boldsymbol{0}$, the quantities $\cG$
take the form $\cG_{l,s} = \frac{g}{\hbar}\, G_{l,s}$ with
\begin{align}
    G_{l,s}
    =
    \int_0^{t_f}dt\,
    \langle\chi_{l,s}(t)|
    W_{\mathrm{int}}
    |\chi_{l=0,+}(t)\rangle,
    \label{esta_G}
\end{align}
which are the time-integrated coupling matrix elements between
the reference channel and the other channels.
We get
\begin{eqnarray}
    f
    =
    1-\frac{g^2}{\hbar^2}
    \sum_{(l,s)\neq(0,+)}|G_{l,s}|^2,
\end{eqnarray}

The components of the control-response vector
$\boldsymbol{K}_{l,s}$ are
\begin{subequations}
\label{esta_K_reference}
\begin{align}
    (\boldsymbol{K}_{l,s})_k
    &=
    \frac{1}{2}
    \int_0^{t_f}dt\,
    \mathcal{L}_k(t)
    \langle\chi_{l,s}(t)|
    \tau_z
    |\chi_{l=0,+}(t)\rangle,
    \\
    (\boldsymbol{K}_{l,s})_{M+k}
    &=
    \frac{1}{2}
    \int_0^{t_f}dt\,
    \mathcal{L}_k(t)
    \langle\chi_{l,s}(t)|
    \tau_x
    |\chi_{l=0,+}(t)\rangle,
\end{align}
\end{subequations}
with $k=1,\ldots,M$. Since the control operators act only on the giant atom,
orthogonality of the waveguide eigenstates gives
\begin{align}
    \boldsymbol{K}_{l,s}=\boldsymbol{0}
    \qquad(l\neq0),
\end{align}
Therefore, it follows
\begin{eqnarray}
\boldsymbol{v}
    &=&
    \frac{g}{\hbar}
    \sum_{(l,s)\neq(0,+)}
    \operatorname{Re}
    \left(
        G_{l,s}^{*}\boldsymbol{K}_{l,s}
    \right) \nonumber\\
&=&
    \label{esta_gradient_channels}
    \frac{g}{\hbar}
    \operatorname{Re}
    \left(
        G_{l=0,-}^{*}\boldsymbol{K}_{l=0,-}
    \right),
\end{eqnarray}

Since the control fields depend linearly on
$\boldsymbol{\lambda}$, we get
\begin{align}
    \frac{\partial^2 \Delta H_g}
    {\partial\lambda_a\partial\lambda_b}
    =0,
    \qquad a,b=1,\ldots,2M,
\end{align}
such that the second-order response matrices entering the general Hessian
expression satisfy
\begin{align}
    (W_{l,s})_{a,b}=0 ,
\end{align}

Substituting $G_{l,s}$ and
$\boldsymbol{K}_{l,s}$ into the general expressions gives
\begin{eqnarray}
    \mathbb{H}_{a,b}
    &=&
    \sum_{(l,s)\neq(0,+)}
    \operatorname{Re}
    \left[
        (\boldsymbol{K}_{l,s}^{*})_a
        (\boldsymbol{K}_{l,s})_b
    \right] \nonumber\\
&=&
    \operatorname{Re}
    \left[
        (\boldsymbol{K}_{l=0,-}^{*})_a
        (\boldsymbol{K}_{l=0,-})_b
    \right] ,
    \label{esta_hessian_channels}
\end{eqnarray}

All these expressions are then substituted into the correction formula from
the preceding subsection,
\begin{align}
    \boldsymbol{\lambda}_{\mathrm{eSTA}}
    =
    -\frac{
        \boldsymbol{v}\|\boldsymbol{v}\|^2
    }{
        \boldsymbol{v}^T\mathbb{H}\boldsymbol{v}
    }.
\end{align}
where $\|\boldsymbol{v}\|$ is the norm.
For a nonzero gradient and negative fidelity curvature along the gradient direction, this expression gives the optimal correction
along the gradient direction within the local quadratic
approximation. Substituting
$\boldsymbol{\lambda}_{\mathrm{eSTA}}$ into the detuning and
Rabi-frequency expressions yields the eSTA control protocol
for our model.

\begin{figure*}[t]
\centering

\includegraphics[width=0.9\linewidth]{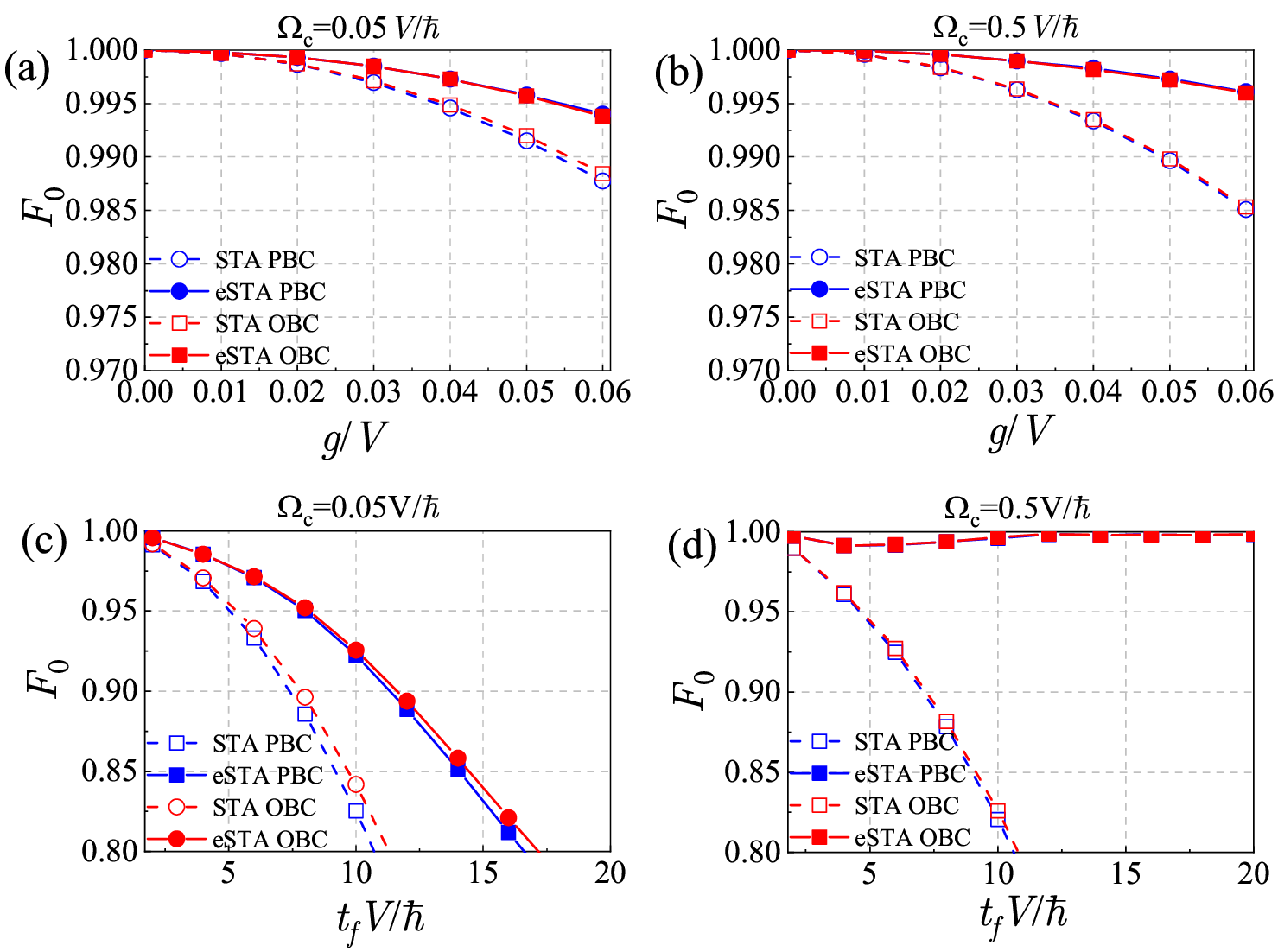}
\caption{Comparison of the fidelity $F_0$ obtained with the STA and eSTA protocols for an atomic waveguide with PBC and OBC. Panels (a) and (b) show $F_0$ with $g$ when the $t_f=2\hbar/V$, while panels (c) and (d) show its dependence on the evolution time $t_f$ when the $g/V=0.05$. The waveguide Rabi frequency is $\Omega_c=0.05V/\hbar$ in (a) and (c), and $\Omega_c=0.5V/\hbar$ in (b) and (d).
\label{fig:STAeSTA0}
}
\end{figure*}

\begin{figure*}[htbp]
\centering
\includegraphics[width=0.9\linewidth]{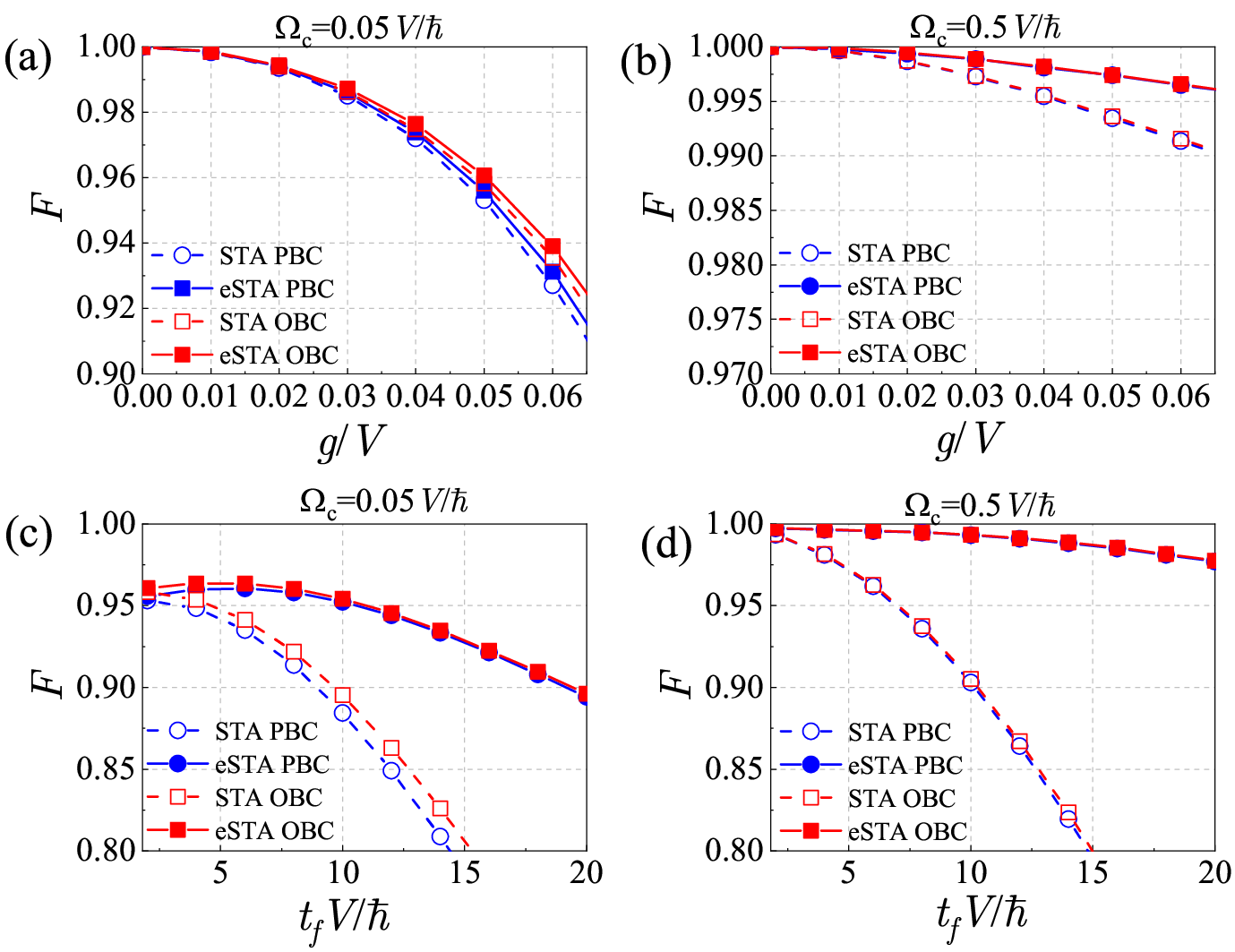}

\caption{Comparison of the fidelity $F$ obtained with the STA and eSTA protocols for an atomic waveguide with periodic (PBC) and open (OBC) boundary conditions. Panels (a) and (b) show $F$ as a function of the coupling strength $g$ at $t_f=2\hbar/V$, while panels (c) and (d) show its dependence on the evolution time $t_f$ at $g=0.05V$. The waveguide Rabi frequency is $\Omega_c=0.05V/\hbar$ in (a) and (c), and $\Omega_c=0.5V/\hbar$ in (b) and (d).
\label{fig:STAeSTA}
}
\end{figure*}

\subsection{Comparison of STA and eSTA Fidelity}

In the first setting, the initial and target states are
$|\Psi_{0,g=0}\rangle$ and $|\Psi_{T,g=0}\rangle$, and the transfer
fidelity is $F_0$, defined in Eq.~\eqref{eq_fidelity0}.
In the second setting, the coupled ground states
$|\Psi_{0,g>0}\rangle$ and $|\Psi_{T,g>0}\rangle$ are used, and the
transfer fidelity is $F$, defined in Eq.~\eqref{full_fidelity}.
Figs.~\ref{fig:STAeSTA0} and~\ref{fig:STAeSTA} respectively compare the variation of final fidelity $F_0$ and $F$ with coupling strength and evolution time under the original STA and eSTA-corrected control functions. In two figures, panels (a) and (b) show the variation of fidelity with coupling strength $g$, corresponding to  $\Omega_c=0.05V/\hbar$ and $\Omega_c=0.5V/\hbar$, respectively. Panels (c) and (d) show the variation of fidelity with evolution time $t_f$, where the coupling strength is fixed at $g=0.05V$.

Panels (a) and (b) of Figs.~\ref{fig:STAeSTA} and~\ref{fig:STAeSTA0} show the fidelity under the original
STA control functions decreases with increasing coupling strength,
and this decrease is more pronounced than under the eSTA-corrected
control functions. These results show that the eSTA correction
can effectively suppress the fidelity loss caused by increasing
the coupling strength. While panels (c) and (d) show that the two protocols give similar fidelities
at short evolution times when the atomic waveguide Rabi frequency is
$\Omega_c=0.05V/\hbar$. However, as $t_f$ increases, the fidelity
decreases faster under the original STA control functions than
under the eSTA-corrected control functions. The difference between
the two protocols therefore becomes more evident at longer
evolution times. When $\Omega_c=0.5V/\hbar$, the improvement
due to the eSTA correction is more significant. The fidelity
under the original STA control functions decreases as $t_f$
increases, while that under the eSTA-corrected control functions
remains close to unity over the considered time range.

These results show that, for the parameters considered,
the eSTA correction constructed from the uncoupled STA reference
dynamics can reduce the fidelity loss during the full coupled
evolution when the initial and target states are chosen as the
ground states of either the uncoupled or the coupled system at the initial and final
times, respectively. The preceding perturbative analysis identifies
the sources of coupling-induced fidelity loss, while the numerical
comparison further demonstrates that the eSTA correction can
suppress this loss.

The results discussed above are obtained for the atomic waveguide
with PBC. Since the boundary conditions affect the eigenstates
and energy gap of the waveguide, we next consider OBC to
examine whether the STA and eSTA protocols retain the features
discussed above.

The effect of the waveguide boundary condition on the fidelities
$F_0$ and $F$ is shown in Figs.~\ref{fig:STAeSTA0}
and \ref{fig:STAeSTA}. Under both boundary conditions, the original
STA protocol gives unit fidelity in the uncoupled system.
For $\Omega_c=0.05V/\hbar$, both $F_0$ and $F$ exhibit a small
dependence on the boundary condition. As shown in
Figs.~\ref{fig:STAeSTA0}(c) and \ref{fig:STAeSTA}(c), the STA
fidelities under OBC are slightly higher than those under PBC
over most of the $t_f$ considered. After the eSTA-corrected
control function is applied, the fidelities under the two
boundary conditions nearly overlap and remain higher than
the corresponding STA results.
When $\Omega_c=0.5V/\hbar$, the difference between the two
boundary conditions becomes much smaller. The STA curves
under PBC and OBC nearly overlap, and the same behavior
is observed for the eSTA curves.

Thus, within the parameter range considered, the boundary
dependence of both $F_0$ and $F$ is more pronounced at the
smaller $\Omega_c$. Changing the boundary condition modifies
the fidelity quantitatively but does not alter the overall
trends of the two control protocols. The eSTA reduces the
fidelity loss under both PBC and OBC.

\section{ \label{conclusion}Conclusion}

In this work, we study enhanced shortcuts to adiabaticity for a giant atom coupled to an interacting atomic waveguide. The weak-coupling analysis shows that both dynamical transitions and coupling-induced changes in the initial and target states contribute to the fidelity loss. The relative importance of these contributions varies with the waveguide driving strength. This explains why state transfer between uncoupled ground states and that between coupled ground states give different fidelities.

The eSTA-corrected control function is obtained from the known STA dynamics of the uncoupled system and improves the final fidelity by adjusting only the giant-atom Rabi frequency and detuning. Numerical calculations show this improvement for both settings of initial and target states, with periodic and open boundary conditions, over the parameter range studied. The original STA protocol for the isolated giant atom can therefore be adapted to improve finite-time state transfer in the interacting system, without the need to construct an exact STA protocol for the full system.

\begin{appendix}

\section{Perturbative analysis of the fidelity loss\label{perturbation}}

To analyze the effect of the coupling between the giant atom and the
atomic waveguide, we write the full Hamiltonian as
\begin{align}
    H_{\mathrm{full}}(t)
    =
    H_0(t)+gW_{\mathrm{int}},
    \label{app_full_H}
\end{align}
where $H_0(t)=H_G(t)+H_c$
is the uncoupled Hamiltonian and $W_{\mathrm{int}}$ denotes the
coupling operator and it can be written as
\begin{align}
    W_{\mathrm{int}}
    =
    S_{+}\tau_{-}
    +
    S_{-}\tau_{+},
    \qquad
    S_{\pm}
    =
    \sigma_\pm^n+\sigma_\pm^m ,
\end{align}

The final fidelity is defined as
\begin{align}
    F(g)
    =
    \left|
    \langle \Phi_T(g)|
    U_{\mathrm{full}}(t_f,0;g)
    |\Phi_0(g)\rangle
    \right|^2,
    \label{app_fidelity}
\end{align}
where $|\Phi_0(g)\rangle$ and $|\Phi_T(g)\rangle$ are the ground states of the full Hamiltonian at $t=0$ and $t=t_f$,
respectively.

In the weak-coupling regime (small $g$), the initial and target states are expanded
as
\begin{align}
    |\Phi_j(g)\rangle
    =
    |\Phi_j^{(0)}\rangle
    +
    g|\Phi_j^{(1)}\rangle
    +
    g^2|\Phi_j^{(2)}\rangle
    +
    O(g^3),
    \;
    j=0,T,
    \label{app_boundary_expansion}
\end{align}
We demand that the state $|\Phi_j(g)\rangle$ is normalised to one for all $g$; this leads to
the conditions $\mbox{Re} \langle\Phi_j^{(0)}|\Phi_j^{(1)}\rangle=0$
and $2\mbox{Re} \langle\Phi_j^{(0)}|\Phi_j^{(2)}\rangle=-\fnormq{\Phi_j^{(1)}}$.

The reference STA protocol satisfies $U^{(0)}(t_f,0)|\Phi_0^{(0)}\rangle
    =|\Phi_T^{(0)}\rangle$
The evolution operator is similarly expanded as
\begin{align}
    U_{\mathrm{full}}(t_f,0;g)
    =
    U^{(0)}(t_f,0)
    +
    gU^{(1)}(t_f,0)\nonumber\\
    +
    g^2 U^{(2)}(t_f,0)
    +
    O(g^3),
    \label{app_U_expansion}
\end{align}
with
\begin{align}
    U^{(1)}(t_f,0)
    =
    -\frac{i}{\hbar}
    \int_0^{t_f}dt\,
    U^{(0)}(t_f,t)
    W_{\mathrm{int}}
    U^{(0)}(t,0),
    \label{app_U1}
\end{align}
and
\begin{align}
\lefteqn{U^{(2)}(t_f,0)
    =} & & \nonumber\\
    & &
    -\frac{1}{\hbar^2}
    \int_0^{t_f}dt \int_0^t ds\,
    U^{(0)}(t_f,t)
    W_{\mathrm{int}}
    U^{(0)}(t,s) W_{\mathrm{int}}
    U^{(0)}(s,0),
    \label{app_U1}
\end{align}

The evolved state at $t_f$ is therefore
\begin{align}
    |\Psi(t_f;g)\rangle
    =
    |\Psi^{(0)}\rangle
    +
    g|\Psi^{(1)}\rangle
    +
    g^2|\Psi^{(2)}\rangle
    +
    O(g^3),
\end{align}
where
\begin{subequations}
\begin{align}
    |\Psi^{(0)}\rangle
    &=
    U^{(0)}(t_f,0)
    |\Phi_0^{(0)}\rangle,
    \\
    |\Psi^{(1)}\rangle
    &=
    U^{(0)}(t_f,0)
|\Phi_0^{(1)}\rangle
    +
    U^{(1)}(t_f,0)
    |\Phi_0^{(0)}\rangle,\\
|\Psi^{(2)}\rangle
    &=
    U^{(0)}(t_f,0)
    |\Phi_0^{(2)}\rangle
    +
    U^{(1)}(t_f,0)
    |\Phi_0^{(1)}\rangle
    \nonumber\\
    &+U^{(2)}(t_f,0)
    |\Phi_0^{(0)}\rangle,
\end{align}
\end{subequations}

We introduce the projector onto the target state,
\begin{align}
    P_T(g)
    &=
    |\Phi_T(g)\rangle
    \langle\Phi_T(g)|, \nonumber \\
     1-F(g)&=\|(1-P_T(g))|\Psi(t_f;g)\rangle\|^2 ,
\end{align}
When we now calculate the fidelity, we get
\begin{eqnarray}
    F(g) &=&1 - g^2 \cR +
    O(g^3),
\end{eqnarray}
where $\cR=\cR_{d} + \cR_{T} + \cR_{0} + \cR_{i}$
and
\begin{eqnarray}
\cR_{d}&=& -\fabsq{\braket{\Phi_T^{(0)}}{U^{(1)}\Phi_0^{(0)}}}\nonumber\\ &&- 2 \mbox{Re} \braket{\Phi_T^{(0)}}{U^{(2)}\Phi_0^{(0)}},
\\
\cR_{T}&=& -\fabsq{\braket{\Phi_T^{(0)}}{\Phi_T^{(1)}}} + \fnormq{\Phi_T^{(1)}},
\\
\cR_{0}&=& -\fabsq{\braket{\Phi_0^{(0)}}{\Phi_0^{(1)}}} + \fnormq{\Phi_0^{(1)}},
\\
\cR_{i}&=& -2 \mbox{Re}\Big(
\braket{\Phi_0^{(0)}}{\Phi_0^{(1)}}\braket{U^{(1)}\Phi_0^{(0)}}{\Phi_T^{(0)}}
\nonumber\\
& & + \braket{\Phi_0^{(0)}}{\Phi_0^{(1)}}\braket{\Phi_T^{(0)}}{\Phi_T^{(1)}}
\nonumber\\
& & + \braket{\Phi_T^{(1)}}{\Phi_T^{(0)}}\braket{U^{(1)}\Phi_0^{(0)}}{\Phi_T^{(0)}}
\nonumber\\
& & + \braket{\Phi_T^{(1)}}{U^{(0)}\Phi_0^{(1)}}
+ \braket{\Phi_T^{(1)}}{U^{(1)}\Phi_0^{(0)}}
\nonumber\\
& & + \braket{\Phi_T^{(0)}}{U^{(1)}\Phi_0^{(1)}}
\Big) ,
\end{eqnarray}
We get also
$g^2 \cR_{T}=1 - \fabsq{\braket{\Phi_{T}^{(0)}}{\Phi_{T}(g)}} + O(g^3)$ \Big($g^2 \cR_{0}=1 - \fabsq{\braket{\Phi_{0}^{(0)}}{\Phi_{0}(g)}} + O(g^3)$ \Big).
Note that the enhanced STA can be related by taking into account only the term $\cR_{d}$ as the initial and target state are still approximated as unperturbed, from there we get
\begin{eqnarray}
\cR_d = \frac{1}{\hbar^2}\sum_{l,s}^{\prime}
    \left|
        G_{l,s}
    \right|^2,
\end{eqnarray}
where the $G_{l,s}$ are defined in the main text.

\section{Evaluation of $\cR_0$, $\cR_T$ and $\cR_i$ in the giant atom-waveguide setup\label{perturbation2}}

The coupling-induced corrections to the initial and target states are
described by the first-order coefficients
\begin{align}
    c_{l,s}(a)
    =
    \frac{
        B_l^+C_{s,-}(a)
        +
        B_l^-C_{s,+}(a)
    }{
        E_0-E_l+\varepsilon_+(a)-\varepsilon_s(a)
    },
    \qquad
    a=0,t_f,
    \label{boundary_coefficient}
\end{align}
where
$C_{s,\pm}(a)
=\langle\phi_s(a)|\tau^\pm|\phi_+(a)\rangle$,
and $\varepsilon_s(a)$ is the energy of $|\phi_s(a)\rangle$ at the endpoint $a$.
At both endpoints, the invariant and $H_G(a)$ share the
same eigenstates, and the reference branch $|\phi_+(a)\rangle$
corresponds to the ground state of $H_G(a)$.
So we get
\begin{eqnarray}
    \cR_0 = \sum_{l,s}^{\prime} \fabsq{c_{l,s} (0)}, \cR_T = \sum_{l,s}^{\prime} \fabsq{c_{l,s} (t_f)} ,
\end{eqnarray}
The coefficient in Eq.~\eqref{boundary_coefficient} is defined
for $(l,s)\neq(0,+)$, assuming a nondegenerate uncoupled
ground state at each endpoint.

The relative phase accumulated by the channel $(l,s)$ with respect to
the reference channel defined in the main text is
\begin{align}
    \theta_{l,s}
    =
    -\frac{(E_l-E_0)t_f}{\hbar}
    +
    \alpha_s(t_f)-\alpha_+(t_f) ,
    \label{relative_phase}
\end{align}
Accordingly, the coupling-induced endpoint contribution is determined
by
\begin{align}
    D_{l,s}^{\mathrm{cou}}
    =
    c_{l,s}(0)e^{i\theta_{l,s}}
    -
    c_{l,s}(t_f) ,
    \label{boundary_amplitude}
\end{align}
We get
\begin{eqnarray}
\cR_{i}
    &=&
    \frac{2}{\hbar}
    \operatorname{Im}
    \left[
    \sum_{l,s}^{\prime}
    \left(D_{l,s}^{\mathrm{cou}}\right)^*
    e^{i\theta_{l,s}}
    G_{l,s}
    \right] \nonumber\\
    & & -
    \sum_{l,s}^{\prime}
    \Big[
    2\operatorname{Re}
    \big(
    c_{l,s}(0)
    [c_{l,s}(t_f)]^*
    e^{i\theta_{l,s}}
    \big)
    \Big].
    \label{app_Rcou_expanded}
\end{eqnarray}

\end{appendix}

\bibliography{reference}

\end{document}